\documentclass[twocolumn,showpacs,preprintnumbers,amsmath,amssymb]{revtex4}
\usepackage{graphicx}% Include figure files
\usepackage{dcolumn}% Align table columns on decimal point
\usepackage{bm}% bold math

\usepackage{times}
\usepackage{float}
\usepackage{flafter}
\usepackage{epsfig}%

\begin{document}

%\preprint{Submitted to Phys. Rev. Lett. 13/08/2002}
%\twocolumn[\hsize\textwidth\columnwidth\hsize\csname@twocolumnfalse\endcsname 
%\title{Do glasses flow like liquids ? }
\title{Shear localization in a model glass}
%\title{Spatial heterogeneities in flowing glassy materials}
\author{F. Varnik$^{(1)}$, L. Bocquet$^{(2)}$, J.-L. Barrat$^{(2)}$, L. Berthier$^{(3)}$}
%\affiliation
\address{(1) CECAM, ENS-Lyon, 46 All\'ee d'Italie, 69007 Lyon, France\\
(2) D\'epartement de Physique des Mat\'eriaux, Universit\'e Lyon I and CNRS, 69622 Villeurbanne Cedex, France\\
(3) Theoretical Physics, Oxford University, 1 Keble Road, Oxford, OX1 3NP, UK}
\date{\today}% It is always \today, today,

\begin{abstract}
Using molecular dynamics simulations, we show that a simple model of a glassy material exhibits the shear localization phenomenon observed in many complex fluids. At low shear rates, the system separates into a fluidized shear-band and an unsheared part. The two bands are characterized by a very different dynamics probed by a local intermediate scattering function. Furthermore, a stick-slip motion is observed at very small shear rates. Our results, which open the possibility of exploring complex rheological behavior using simulations, are compared to recent experiments on various soft glasses.
\end{abstract}

% PACS, the Physics and Astronomy Classification Scheme:
%%05.70.Ln   Nonequilibrium and irreversible thermodynamics (see also 82.40.Bj Oscillations, chaos, and bifurcations in physical chemistry and chemical physics)
%%64.70.Pf   Glass transitions  
%%83.60.Fg Shear rate dependent viscosity
\pacs{64.70.Pf,05.70.Ln,83.60.Df,83.60.Fg}

\maketitle

%\keywords{Suggested keywords}%Use showkeys class option
%if keyword display desired

\vskip2pc

%\section{\label{sec:level1}Introduction}

Shear localization is a commonly observed phenomenon in the rheology of complex fluids. Over some range of shear rates, a fluid undergoing simple shear flow in, say, a Couette cell tends to separate into bands parallel to the flow direction, with high shear rate regions coexisting with smaller shear rate regions.  In some cases~\cite{olmsted}, this shear-banding phenomenon can be understood in terms of underlying structural changes in the fluid, analogous to a first order phase transition. In other systems, however, no such changes are evident, and coexistence appears between a completely steady region (zero shear rate)  and a sheared, fluid region. This second type of behavior has been observed \cite{Chen,Pignon::JRheo40::1996,Losert::PRL85::2000,Debregeas::PRL87::2001,Coussot-Raynaud-et-al::PRL88::2002}, in particular in systems of the so-called `soft glass' type~\cite{Sollich}. Such systems include dense colloidal pastes, granular materials, emulsions, etc. Their rheological behavior is essentially determined by the competition between an intrinsic slow dynamics and the acceleration caused by  the external flow~\cite{Sollich,BB::PRE61::2000,BB2}. The large time scales inherent to the glassy state manifest themselves in the non-linear character of the rheological properties as a function of the shear rate $\gamma$: existence of a yield stress (the system does no
flow until the stress $\sigma$ exceeds a threshold value $\sigma_{\text y}$) and non-linear flow curves $\sigma=\sigma(\gamma)$, leading to a shear rate dependent viscosity.

The observation of strong heterogeneities in the flow of such systems suggests that a global flow curve is not sufficient to fully characterize the flow behavior. Indeed, a simple shear-thinning behavior would in general imply homogeneous shear flow in a planar Couette cell, since the shear stress is constant across the cell. More generally, it remains to clarify if these observations in systems with very different microscopic interactions are intrinsic to glassy dynamics, that is, if a generic scenario for inhomogeneous shear flow can be proposed, as attempted in several recent studies~\cite{Dhont,Picard}.

In order to investigate this issue, we performed molecular dynamics  simulations of  a simple and generic glass forming system,  consisting of a binary Lennard-Jones mixture. Since our work is done on a simple liquid, it can be seen as a link between experiments on complex materials and pure phenomenology. %Using molecular simulation, despite the obvious limitations on time and length scales, is a way to bypass some of the difficulties encountered in experiments, e.g. the determination of velocity profiles or the determination of flow curves under strictly  homogeneous shear conditions.

The present model system has been extensively studied~\cite{BB::PRE61::2000,BB2,Kob-Andersen}. It is known to exhibit, in the bulk state,  a computer glass transition (in the sense that the relaxation time becomes larger than typical simulation times) at a temperature $T_{\text c} \simeq 0.435$ for a 80:20 mixture at a density $\rho=1.2$ (in Lennard-Jones units~\cite{Kob-Andersen}). All our simulation parameters are chosen equal to those in Refs.~\cite{BB2,Kob-Andersen}. Since our aim is to analyse possible flow heterogeneities, we do not impose a constant velocity gradient over the system as was done in Ref.~\cite{BB2}, where a homogeneous shear flow was imposed through the use of Lee-Edwards boundary conditions. Rather we confine the system between two solid walls, which will be driven at constant velocity. By doing so, we mimic an experimental shear cell, without imposing an homogeneous flow.

We first equilibrate a large simulation box with periodic boundary conditions in all directions, at $T=0.5$. The system is then quenched to a temperature below $T_{\text c}$. Most of the results discussed in this work correspond to $T=0.2$. At this temperature, the structural relaxation times are orders of magnitude larger than at $T_{\text c}$. On the time scale of computer simulations, the system is in a glassy state, in which its properties slowly evolve with time (aging).  After a  time of $t=4.10^4$ [$2.10^6$ MD steps], we create  2 parallel solid boundaries by freezing all the particles outside  two parallel $xy$-planes at positions $z_{\text {wall}}= \pm L_z/2$.   For each computer experiment, 10 independent samples (each containing 4800 particles) are prepared using this procedure. The  cell containing the fluid particles has dimensions  $L_x=L_y=10$ and  $L_z=40$.

An overall shear rate is imposed by moving in the $x$-direction all the atoms of, say, the left wall ($z_{\text {wall}}=-20$) with a strictly constant velocity of $U_{\text {wall}}$. This defines the total shear rate  $\gamma_{\text {tot}}=U_{\text {wall}}/L_z$. Velocity profiles are recorded discarding transients (of typical duration $1/\gamma$) associated with the start of the shear motion. In order to remove the viscous heat created by the shear,  the system is coupled to a heat bath. For this purpose, we divide the system into parallel layers of thickness $dz=0.25$ and rescale (once every 10 integration steps) the $y$-component of the particle velocities within the layer, so as to  impose the desired temperature $T$. Such a local treatment is necessary to keep a homogeneous temperature profile when flow profiles are heterogeneous. To check for a possible influence of the thermostat, we compared, for an extremely low shear rate ($\gamma_{\text {tot}}= U_{\text{wall}}/L_z=10^{-5}$), these results with the output of a simulation where the inner part of the system was unperturbed and the walls were thermostatted instead. Both methods give identical results, indicating that the results are  not affected by the thermostat.
\begin{figure}
\hspace*{-1mm}\epsfig{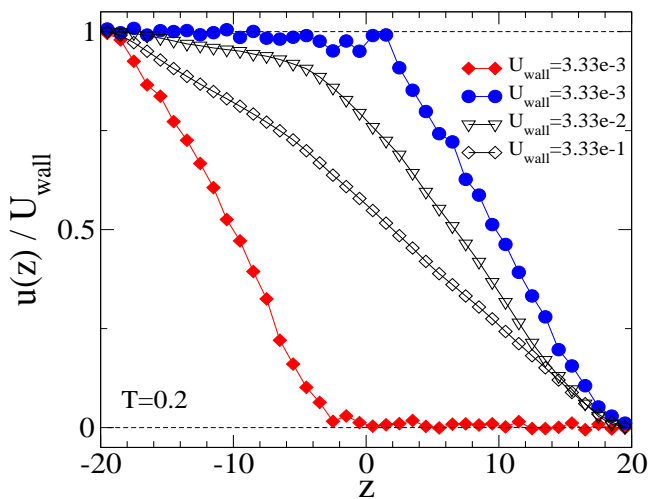}
\caption{\label{fig:fig1} Filled symbols: rescaled velocity profiles, $u(z)/U_{\text{wall}}$, from 2 independent simulation runs. In both cases, the left wall is moved with a constant velocity $U_{\text {wall}}=3.33\times 10^{-3}$ ($\gamma_{\text{tot}}=0.83\times 10^{-4}$). Due to Galilean invariance, the sheared region may be located at either the moving or immobile wall. Open symbols: rescaled velocity profiles obtained at lower wall velocities $U_{\text {wall}}=3.33\times 10^{-2} $ and $3.33\times 10^{-1}$ corresponding to overall shear rates of $\gamma_{\text{tot}}=0.83\times 10^{-3}$ and $0.83\times 10^{-2}$. Note that the local shear rate of the sheared region is smaller at smaller $U_{\text{wall}}$.}
\end{figure}

Our main observation is reported in Fig.~\ref{fig:fig1}. We find that, at low overall shear rates $\gamma_{\text {tot}}$, the system separates into a homogeneously sheared band and a part which is essentially unsheared. Note that, the velocity profiles are  not symmetric with respect to the midplane. Rather, the shear band is localized close to one of the walls, so that a single interface between sheared and steady region is formed. The symmetry between the two walls, which is a result of Galilean invariance, is restored only on average, the shear band occurring equally likely on both sides of the simulation cell (see Fig.~\ref{fig:fig1}). A similar behavior has been observed in various experiments~\cite{Chen,Pignon::JRheo40::1996,Losert::PRL85::2000,Debregeas::PRL87::2001,Coussot-Raynaud-et-al::PRL88::2002}.

In some cases, we observed oscillations of the velocity profile between the two mentioned solutions. This effect is possibly related to the finite size of the simulation box leading to a finite probability for the band to oscillate. In contrast, this probability would be zero for (very large) experimental systems, thus stabilizing one of the solutions. A discussion of this interesting aspect is beyond the scope of this paper, and we postpone it to future work.

As shown in Fig.~\ref{fig:fig1}, the thickness $h$ of the sheared region depends on the wall velocity $U_{\text {wall}}$. For very small $U_{\text {wall}}$, $h$ is of the order of a few atomic diameters and varies only slightly with the wall velocity. As a consequence, the shear rate inside the sheared region, of order $U_{\text {wall}}/h$, decreases as $U_{\text {wall}}$ is reduced. As $U_{\text{wall}}$ is further increased, $h$ increases to reach the full slab size, $h=L_z$, at a given wall velocity, $U_c$. For $U_{\text {wall}}>U_c$ the velocity profile is therefore linear, as is oberved in simple fluids. These findings are in qualitative agreement with experimental results on clay suspensions~\cite{Pignon::JRheo40::1996}. The rather slow variation of $h$ with respect to a change of $\gamma_{\text tot}$ at small wall velocities should, however, be contrasted to reports in~\cite{Coussot-Raynaud-et-al::PRL88::2002}. However, the small size of our system compared to the width of interfacial regions makes a more quantitative analysis rather difficult for the moment.

At a given global shear rate $\gamma_{\text {tot}}=10^{-4} (U_{\text{wall}}=0.004)$, we  also studied the influence of temperature on shear localization. We find that increasing the temperature is qualitatively similar to increasing the global shear rate: the thickness of the sheared region is an increasing function of the temperature. For example, $h \approx 15$ [see Fig.~\ref{fig:fig1}] at $T=0.2$ whereas $h\approx 25$ at $T=0.4$ and $h\approx L_z(=40)$ at $T=0.5$.

\begin{figure}
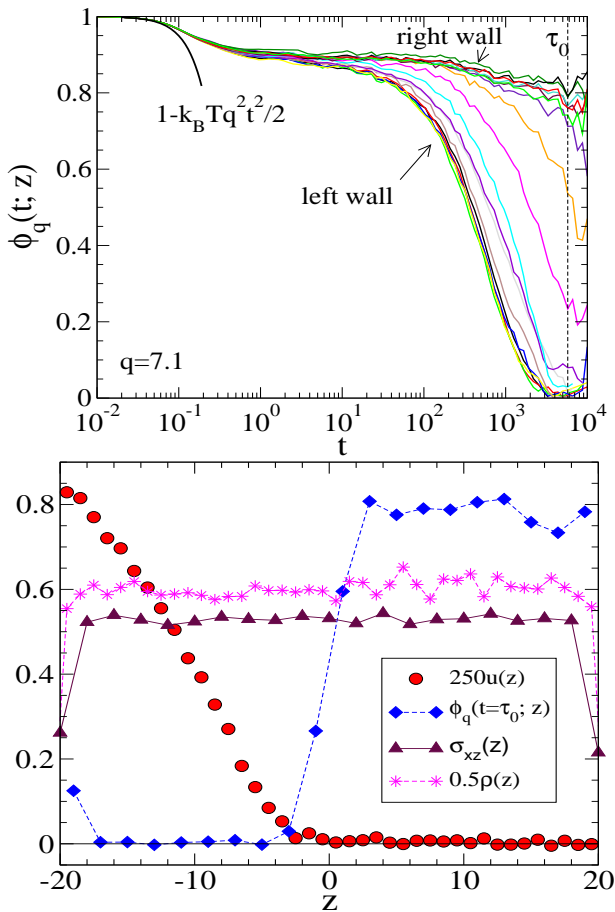

\hspace*{-2mm}\epsfig{file=fig2a.eps, height=60mm, width=80mm,clip=, angle=0, silent=} \hspace*{0mm}\epsfig{file=fig2b.eps,height=60mm, width=80mm, clip=, angle=0, silent=}
\caption{\label{fig:fig2}
Upper panel: Intermediate scattering function, $\phi_{q}(t; z)$, computed within layers of thickness $dz=2$. From the bottom to the top, $z=-17,-15,\cdots,15,17$. The temperature is $T=0.2$, and $\gamma_{\text{tot}} = 0.83\times 10^{-4}$. The vertical dashed line marks the time $\tau_0=5754 \approx 0.5 / \gamma_{\text{tot}}$ [this choice for $\tau_0$ is due to larger statistical noise for $t>5754$]. Lower panel: $\phi_q(\tau_0;z)$ (connected diamonds). Connected circles depict the corresponding velocity profile. Connected triangles shows the local  stress $\sigma(z)$ and connected stars the local density profile $\rho(z)$.}
\end{figure}

We find no obvious structural differences between the two regions of the sample. As shown in Fig.~\ref{fig:fig2} (lower panel),  static properties like the density profile, shear stress, and normal pressure are found to be constant across the film~\cite{shearstress}.

The distinction between the two bands is therefore purely dynamical. This can be seen for instance on the layer-resolved intermediate scattering function, $\phi_{q}(t; z)= \sum_i \left< \exp[ {\text i} q_y(y_i(t)-y_i(0))] \; \delta \left(z_i - z \right) \right>$. We choose $q_y=7.1$, which  corresponds to  the first  maximum  of the static structure factor. The upper panel of Fig.~\ref{fig:fig2} depicts $\phi_q(t;z)$ obtained from a simulation with $\gamma_{\text {tot}}=0.83\times 10^{-4}$. One may first note the existence of two limiting behaviors of the correlation function, corresponding to the sheared and jammed regions, with a rather rapid change from one behavior to the other within a few layers. While in the jammed region (close to the right wall in Fig.~\ref{fig:fig2}) $\phi_q(t;z)$ barely relaxes, the correlation function in the sheared region (close to the left wall)  exhibits a two-step relaxation as observed in homogeneously sheared systems. This observation reflects the acceleration of the structural relaxation due to the flow~\cite{BB2}. In the jammed region, the system behaves as a glassy solid, and  $\phi_q(t;z)$ does not relax to zero on the simulation time scale.

A way to quantify the relation between the structural
 relaxation and the variation of the shear rate across the system is to
look at the quantity $\phi_q(\tau_0;z)$, where $\tau_0$ is of order of
$1/\gamma_{\text {tot}}$. This quantity reflects the way the system
has relaxed on the time scale imposed by the global shear rate.
Results for $\phi_q(\tau_0; z)$ are shown in the lower panel of
Fig.~\ref{fig:fig2}. The change of the velocity profile, when
going from the sheared towards the unsheared region, is
accompanied by a sharp jump in $\phi_q(\tau_0;z)$ at the interface
between these two regions. The profile of $\phi_q(\tau_0; z)$ is
very similar to that of an order parameter across an
interface~\cite{harrowell}. Other order parameters could be
proposed to characterize the local dynamics: for example, the
local relaxation time of $\phi_q(t; z)$, defined as the time after
which $\phi_q(t; z)$ goes below a fraction of unity or the local
diffusion coefficient parallel to the walls. All these definitions
lead to a two-phase picture, with well defined and spatially
constant local characteristics in both phases.

\begin{figure}[htb]
\hspace*{-4mm}\epsfig{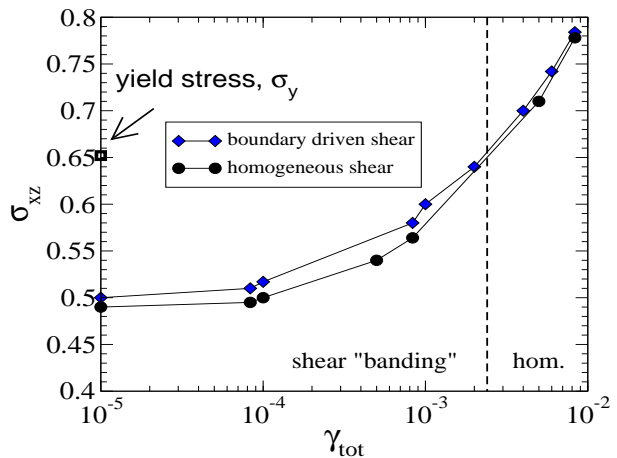} \caption{\label{fig:fig3} Diamonds:
$\sigma$ versus  $\gamma$ under homogeneous flow conditions at
$T=0.2$. Circles: $\sigma$ versus $\gamma_{\text {tot}}$ in the
boundary-driven shear flow. The vertical dashed line is an
estimate of the global shear rate below which shear localization
is expected.}
\end{figure}

In Fig.~\ref{fig:fig3}, we summarize the results by plotting
the $\sigma(\gamma)$ flow curve
of the model at $T=0.2$. In this
 figure, we indicate the flow curve obtained for a system
 undergoing homogeneous shear flow, taken from
 Ref.~\cite{BB2}. We also show the points obtained for the
 same system driven by the boundaries, as described in the present
 paper. Finally, we indicate the static yield stress of the system, $\sigma_{\text y}$.

To obtain $\sigma_{\text y}$, we apply a small tangential force, $F_{\text T}$, acting on the left wall (treated as a rigid object with overdamped dynamics) for a certain amount of time, during which the velocity profile, $u(z)$, is sampled. The force on the wall is then slightly increased for a new measurement before going over to the next higher value. The static yield stress is then defined as the smallest force  (per unit area) for which the average streaming velocity in the left half of the system, $u_{\text{av, left}}=\int_{-20}^{0} u(z)dz/20$, exeeds $u_{\text min}$, a minimum value, a few times larger than the typical statistical error of $u_{\text{av, left}}$. We empirically find that, for a measurement time of $4.10^{3}$ [$5.10^{4}$MD steps], $u_{\text min}=4.10^{-4}$ is a reasonable choice. The result on $\sigma_{\text y}$ is then averaged over $10$ independent runs. Again, for each initial configuration, the system is first equilibrated at $T=0.5$ before being quenched to a temperature below $T_{\text c}=0.435$. This ensures the equivalence of all individual runs (no history dependence). The system is then propagated with $F_{\text T}=0$ for a time $t_{\text w}$ before the very first increment of the tangential force. At $T=0.2$ and using an increment of $dF_{\text T}=0.02$ (once in $5.10^{4}$MD steps) we have obtained $\sigma_{\text y}=0.596 \pm 0.022,\; 0.658 \pm 0.009$ and $0.652 \pm 0.015$ corresponding to waiting times of $t_{\text w}=10^{3}, 4.10^{3}$ and  $4.10^{4}$ respectively. Thus, already at a waiting time of $t_{\text w}=4.10^3$, aging effects on $\sigma_{\text y}$ are negligible at the temperature studied. Note that, due to aging effects, $\sigma_{\text y}$ might also depend on the speed at which the threshold value of the tangential force is reached. Therefore, we have determined $\sigma_{\text y}$ for a higher value of $dF_{\text T}=0.05$ in the case of $T=0.2$ and $t_{\text w}=4.10^{4}$. This gives $\sigma_{\text y}=0.66\pm 0.015$ which agrees well with $0.652 \pm 0.015$ within the error bars. %A waiting time of $t_{\text w}=4.10^4$ is therefore sufficiently long to ensure the independence of $\sigma_{\text y}$ of the increment rate of $F_{\text T}$.

It can be seen in Fig.~\ref{fig:fig3} that $\sigma_{\text y}>\sigma(\gamma_{\text {tot}}\rightarrow 0)$. Therefore, shear-banding can be expected in the region limited by the vertical dotted line, which corresponds to $\sigma(\gamma_{\text {tot}}) < \sigma_{\text y}$. Once the yield stress $\sigma_{\text y}$ is added to the flow curve, the shear rate becomes multivalued in a range of shear stress, a situation encountered in several complex fluids~\cite{olmsted}. This is the very origin of the shear-banding we observe. As a consequence, this phenomenon should be generic for many soft glassy materials.

\begin{figure}[htb]
\hspace*{-4mm}\epsfig{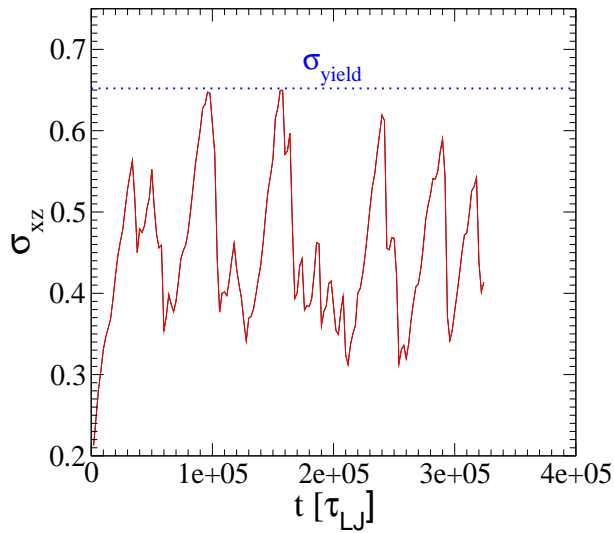} \caption{\label{fig:fig4} Shear stress
versus time for $\gamma_{\text {tot}}=0.83\times 10^{-6}$ at
$T=0.2$. The stress rises up  to a value close to $\sigma_{\text
y}\simeq 0.65$,  before suddenly dropping to a value smaller than
the one obtained in an homogeneous flow ($\sigma_{\text xz}\simeq 0.4)$.}
\end{figure}

Finally, in  the very low shear rate region,  we observe a time dependence of the shear stress characteristic of stick-slip behavior, as demonstrated in Fig.~\ref{fig:fig4}. This is reminiscent to what is observed in friction simulations~\cite{robbins}. Due to numerical limitations, the limit between continuum sliding and stick-slip behavior could not be precisely located. Qualitatively, this behavior is obtained when the thickness of the sheared layer $h$ becomes of the order of a few particle diameters, which also corresponds to the width of the interface separating the sheared and jammed regions.% (recall the lower panel of Fig.~\ref{fig:fig2}).

Our results lead to  the following picture. (i) For global shear rates $\gamma_{\text {tot}}$ smaller than a critical value, $\gamma_c=U_c/L_z$, the system separates spatially into two
regions, one with a finite, approximatively uniform, shear rate,
the other being jammed; (ii) when the total shear rate is
increased, the thickness of the sheared region increases; (iii)
for $\gamma_{\text {tot}}>\gamma_c$, $h$ reaches the thickness of
the slab, so that the flow is homogeneous with a linear velocity
profile; (iv) at very small total shear rates, a stick-slip
phenomenon is obtained; (v) the flow profile closely follows the
local dynamics of the material, the strongly sheared region
corresponding to a rapid structural relaxation, the jammed region
to a glassy solid.

The qualitative agreement of these phenomena with experimental
observations in very different systems is remarkable:
Ref.~\cite{Pignon::JRheo40::1996} describes explicitly the same
scenario for the flow behavior of a Laponite clay suspension,
including the stick-slip at very small shear rates. A more
quantitative study of the flow profiles has been performed in
Ref.~\cite{Coussot-Raynaud-et-al::PRL88::2002} using a Magnetic
Resonance Imaging technique, confirming the existence of well
defined jammed and flow regions, whose relative width depends on
the global shear rate. In contrast to molecular simulations, an
experimental determination of  velocity profiles together with a
local probe of the dynamics is difficult, which makes the
observation of such effects in simulations particularly promising.

Our results do also  constrain possible phenomenological descriptions of the flow scenario. It is indeed important to remark  that models relating the local shear stress to the local shear rate, as $\sigma=\sigma_c+\alpha \gamma^n$ ($\alpha$ being a constant and $n=1$ corresponding to  Bingham fluids, and $n>1$ to Herschel-Buckley fluids), are unable to account for these results. As emphasized above, the shear stress is constant throughout the Couette cell, so that any of the afore-mentioned models would predict a constant shear rate in the cell, in contrast to the present results.  It appears therefore necessary to include explicitly non-local terms in  phenomenological descriptions~\cite{olmsted,Dhont}. The existence of a close connection between velocity profile and local dynamics supports some of the assumptions of recent approaches treating the local `fluidity' of the system as an `order parameter'~\cite{Picard}. More detailed simulations, with systematic investigation of size effects, should allow a direct comparison with the predictions of such models.

F.V. thanks the Deutsche Forschungsgemeinschaft (DFG) and the CECAM for financial support. L.B. is supported by the EPSRC. Generous grants of simulation time by the computer centers at the University of Mainz (ZDV) and at ``l'Ecole Normale Sup\'erieure de Lyon'' (PSMN) are also acknowledged.

%\end{thebibliography}

\end{document}